\documentclass[twocolumn,prl,showpacs,amsmath,amstex,amssymb,mathfonts,superscriptaddress]{revtex4-1}
\pdfoutput=1
\usepackage{amsthm,color,amsfonts,graphicx,verbatim}
\usepackage{amsmath}
\usepackage{amssymb}
\usepackage{amsthm}
\usepackage{amsfonts}
\usepackage{dsfont}
\usepackage{listings}
\usepackage{enumerate}
\usepackage{latexsym}
\usepackage{psfrag}
\usepackage{bm}
\usepackage{graphicx}
\usepackage{color}
\usepackage{hyperref}
\hypersetup{
    bookmarks=true,         
    unicode=false,          
    pdftoolbar=true,        
    pdfmenubar=true,        
    pdffitwindow=false,     
    pdfstartview={FitH},    
    pdftitle={},    
    pdfauthor={},     
    pdfsubject={},   
    pdfcreator={},   
    pdfproducer={}, 
    pdfkeywords={keyword1} {key2} {key3}, 
    pdfnewwindow=true,      
    colorlinks=true,       
    linkcolor=black,          
    citecolor=blue,        
    filecolor=magenta,      
    urlcolor=blue           
}

\newcommand{\be}{\begin{equation}}
\newcommand{\ee}{\end{equation}}
\newcommand{\bea}{\begin{eqnarray}}
\newcommand{\eea}{\end{eqnarray}}

\newcommand{\la}{\langle}
\newcommand{\ra}{\rangle}

\renewcommand{\phi}{\varphi}
\renewcommand{\epsilon}{\varepsilon}
\renewcommand{\vec}[1]{{\bf #1}}

\begin{document}

\title{Constructing local integrals of motion in the many-body localized phase}

\author{Anushya Chandran}
\affiliation{Perimeter Institute for Theoretical Physics, Waterloo, ON N2L 2Y5, Canada}
\author{Isaac H. Kim}
\affiliation{Perimeter Institute for Theoretical Physics, Waterloo, ON N2L 2Y5, Canada}
\author{Guifre Vidal}
\affiliation{Perimeter Institute for Theoretical Physics, Waterloo, ON N2L 2Y5, Canada}
\author{Dmitry A. Abanin}
\affiliation{Perimeter Institute for Theoretical Physics, Waterloo, ON N2L 2Y5, Canada}
\affiliation{Institute for Quantum Computing, Waterloo, ON N2L 3G1, Canada}

\date{\today}
\begin{abstract}

Many-body localization provides a generic mechanism of ergodicity breaking in quantum systems. In contrast to conventional ergodic systems, many-body localized (MBL) systems are characterized by extensively many local integrals of motion (LIOM), which underlie the absence of transport and thermalization in these systems. Here we report a physically motivated construction of local integrals of motion in the MBL phase. We show that any local operator (e.g., a local particle number or a spin flip operator), evolved with the system's Hamiltonian and averaged over time, becomes a LIOM in the MBL phase. Such operators have a clear physical meaning, describing the response of the MBL system to a local perturbation. In particular, when a local operator represents a density of some globally conserved quantity, the corresponding LIOM describes how this conserved quantity propagates through the MBL phase. Being uniquely defined and experimentally measurable, these LIOMs provide a natural tool for characterizing the properties of the MBL phase, both in experiments and numerical simulations. We demonstrate the latter by numerically constructing an extensive set of LIOMs in the MBL phase of a disordered spin chain model. We show that the resulting LIOMs are quasi-local, and use their decay to extract the localization length and establish the location of the transition between the MBL and ergodic phases.

\end{abstract}
\maketitle

\section{Introduction}

A central postulate of statistical mechanics is that quantum systems of many interacting particles, prepared in physical initial states, reach local thermal equilibrium as a result of Hamiltonian evolution. Such systems are called ergodic. The dynamics of ergodic systems is constrained only by a handful of extensive conserved quantities -- energy, momentum, and particle number being familiar examples -- and their subsystems thermalize to Gibbs ensembles set by the values of these conserved quantities. Microscopically, thermalization is believed to follow from the eigenstate thermalization hypothesis (ETH), which conjectures that individual many-body eigenstates in ergodic systems locally appear thermal~\cite{Deutsch91,Srednicki94,Rigol08}. Thus, eigenstates with the same values of the globally conserved quantities are locally indistinguishable. 

While a majority of many-body systems (e.g., a weakly interacting Fermi gas) are empirically known to thermalize, there is mounting evidence that a large class of disordered systems avoid thermalization via a mechanism akin to Anderson localization~\cite{Anderson58} in the Hilbert space~\cite{Gornyi05,Basko06,OganesyanHuse07}. In such {\it many-body localized} (MBL) systems, the laws of statistical mechanics do not apply. 

In contrast to ergodic systems, MBL systems are characterized by extensively many {\it local} conservation laws~\cite{Serbyn13-1,Serbyn13-2,HuseOganesyan13}. This property underlies the breakdown of ETH -- two eigenstates with the same total energy and particle number generally have different values of local integrals of motion, and therefore are locally distinguishable; thus, the subsystems cannot be described by the same Gibbs ensemble. Further, an MBL system retains memory of its initial state during quantum evolution, because the values of local integrals of motion cannot change. 

In addition to providing a fundamental example of a system where statistical mechanics breaks down, the MBL phase can be used for storage and manipulation of quantum information by preparing superpositions of states with different values of local integrals of motion (LIOM). The dynamics in the MBL phase is limited to slow dephasing between different parts of the system~\cite{Serbyn13-1,Serbyn13-2,HuseOganesyan13,Vosk13}, which can be suppressed by preparing particular initial states~\cite{Serbyn13-1}. Moreover, quantum information can be recovered using spin-echo techniques~\cite{Serbyn14-1}. It was also recently argued that MBL can preserve symmetry-breaking, as well as topological and symmetry-protected topological order, at non-zero temperature even when statistical mechanics forbids it~\cite{Huse13,BauerNayak13,Pekker14,Bahri13,Chandran14,Kjall14}; this raises an interesting possibility of topological quantum computation in MBL systems.

The choice of LIOMs in the MBL phase is highly arbitrary. For example, a linear combination or a product of two or more LIOMs is also a LIOM. Which principle should be used to construct LIOMs in MBL systems? Previously, LIOMs have been viewed formally, as quasi-local operators which label eigenstates. In Refs.~\cite{Serbyn13-1,HuseOganesyan13}, it was argued that, in principle, for a finite-size system in which all states are MBL, one can define a minimal complete set of LIOMs, such that their eigenvalues uniquely label eigenstates. However, there are exponentially many ways to define a minimal complete set of LIOMs, and therefore its construction is practically challenging and is currently lacking. 

In this Article, we introduce a new extensive set of {\it measurable} LIOMs in systems which are MBL at all energy densities. 
We show that any local operator ${\mathcal O}$, time-evolved with the Hamiltonian and averaged over time (which we denote by $\bar{\mathcal O}$) becomes a local integral of motion in the MBL phase. LIOMs constructed in this way 
describe the state of an MBL system at long times, following a local perturbation by the operator ${\mathcal O}$. Thus, $\bar{\mathcal O}$ is physically measurable. The LIOMs have a particularly intuitive meaning for the cases when ${\mathcal O}$ represents a conserved density, e.g. when ${\mathcal O}$ is a local particle number or an energy density operator. The LIOM $\bar{\mathcal O}$ determines how far the corresponding conserved quantity, initially perturbed locally, propagates through the MBL system. In contrast to previous constructions~\cite{Serbyn13-1,HuseOganesyan13}, LIOMs introduced in this paper are unique (for any local operator the corresponding LIOM is uniquely defined) and can be obtained numerically in various models of many-body localization. 

We also note that two recent works have perturbatively constructed LIOMs~\cite{Imbrie14,Muller14} of the kind considered in Refs.~\cite{Serbyn13-1,HuseOganesyan13}: Ref.~\cite{Imbrie14} proves the existence of LIOMs near the diagonal limit of a 1D spin model while Ref.~\cite{Muller14} follows the approximate perturbative arguments of Basko, {\it et al.}~\cite{Basko06} near Anderson localized free fermions. In order to explore the characteristics of LIOMs more generally, such as the decomposition in terms of physical operators and localization length, it is necessary to have a physically motivated non-perturbative construction. This is the goal of our work.

We demonstrate the power of our approach by explicitly constructing extensive sets of LIOMs in the MBL phase of a disordered spin model. We explore the properties of the resulting LIOMs, including their decomposition in terms of physical operators. We find that in the MBL phase, integrals of motion (IOMs) are local and have exponentially decaying tails, which we use to extract the localization length. We also define an order parameter for the MBL phase using the projection of the time-averaged conserved density on the site about which the IOM is localized. In the MBL phase, the order parameter is non-zero in the thermodynamic limit. On the other hand, we show that in the delocalized phase the order parameter approaches zero, the localization length diverges and the IOMs are non-local in the thermodynamic limit.


\section{Many-body localization and local integrals of motion}

We start by reviewing the properties of the MBL phase, following Refs.~\cite{Serbyn13-1,HuseOganesyan13}. We consider systems in which all eigenstates are MBL. It was conjectured~\cite{Serbyn13-1} that the key property of the MBL phase is that the system's Hamiltonian can be diagonalized by a sequence of quasi-local unitary transformations.  This is closely related to the small amount of entanglement between remote degrees of freedom in MBL eigenstates. This conjecture has several implications. First, all MBL eigenstates, except for an exponentially small fraction, have an area-law entanglement entropy, typical of ground states in gapped systems; this is supported by numerical simulations of spin chains~\cite{BauerNayak13,Serbyn13-1}. 
A second implication, which is of interest to us, is the existence of a complete set of LIOMs. Third, as we briefly describe at the end of the Section, this conjecture allows one to understand dynamical properties of the MBL phase. We also note that numerical, perturbative and rigorous studies support the conjecture formulated above~\cite{Serbyn13-1,Muller14,Imbrie14}. 

To be specific, consider a 1D XXZ spin chain with nearest-neighbour interactions and random magnetic field in the $z$ direction, described by the Hamiltonian
\be\label{eq:XXZ}
H=J_x \sum_i (\sigma_i^x \sigma_{i+1}^x+\sigma_i^y \sigma_{i+1}^y)+ J_z\sum_i \sigma_i^z \sigma_{i+1}^z+\sum_i h_i \sigma_i^z, 
\ee 
where $h_i$ on each site is randomly distributed in an interval $[-W;W]$. Numerical studies ~\cite{PalHuse,Prosen08,Luca13,Moore12,Reichman14} indicate that this model exhibits an MBL phase at sufficiently strong disorder. As disorder strength is decreased, the system undergoes a phase transition into an ergodic phase, in which the eigenstate thermalization hypothesis holds~\cite{PalHuse}. 

Let us consider a product state basis, $|s_1 s_2 ...s_N\ra=|s_1\ra\otimes |s_2\ra\otimes \dots \otimes |s_N\ra$, $s_i=\uparrow,\downarrow$, which is an eigenbasis for all $\sigma_i^z$ operators ($N$ is the number of spins). Then, there exists a quasi-local unitary transformation $U$, which relates the eigenstates of Hamiltonian (\ref{eq:XXZ}) in the MBL phase to product states~\cite{Serbyn13-1} (see also Ref.~\cite{Imbrie14}, which considers a different 1D spin model). In terms of this unitary transformation, LIOMs can be defined as follows:
\be\label{eq:tau}
\tau_i^z=U \sigma_i^z U^\dagger. 
\ee
$\tau_i^z$ can be viewed as a pseudospin-$1/2$ operator, which is diagonal in the basis of eigenstates of $H$, with an eigenvalue given by $s_i$; thus, it is an integral of motion. Moreover, since it is obtained by acting with a quasi-local unitary on a local operator $\sigma_i^z$, it is quasi-local -- that is, its action on remote spins is exponentially close to unity. 
Thus, the operator $\tau_i^z$ can be approximated by a sum of $\sigma_j^\alpha$ operators and their products, acting on spins within distance $k$ from site $i$, with an exponential accuracy in distance $|i-k|$. 

Operators $\tau_i^z$ are independent and commuting; they form a complete set of LIOMs in the sense that specifying their eigenvalues uniquely specifies an eigenstate. In order to obtain the full set of operators, which can be used to generate any state in the Hilbert space, define ``dressed" $x,y$ pseudo-spin operators as follows:
\be\label{eq:taux}
\tau_i^x=U \sigma_i^x U^\dagger, \,\, \tau_i^y=U \sigma_i^y U^\dagger.  
\ee
The $\tau_i^\alpha$ operators inherit commutation relations of the $\sigma_i^\alpha$ operators, that is, $[\tau_i^\alpha,\tau_j^\beta]=2i \delta_{ij}\epsilon^{\alpha \beta\gamma}\tau_i^\gamma$. Operators $\tau_i^\alpha$, $\alpha=0,x,y,z$, and their various products form a basis in operator space, and any operator can be expanded in such a basis. 

The original Hamiltonian (\ref{eq:XXZ}) has a particularly simple representation in terms of $\tau$-operators. Since it commutes with every $\tau_i^z$, it can only involve $\tau_i^z$ operators~\cite{HuseOganesyan13,Serbyn13-1,Serbyn13-2}: 
\be\label{eq:Htau}
H=\sum_i \tilde H_i \tau_i^z+\sum_{ij} J_{ij} \tau_i^z \tau_j^z + \sum_{ijk} J_{ijk} \tau_i^z \tau_j^z \tau_k^z+\dots
\ee
The couplings between remote clusters of pseudospins decay exponentially with the separation between them: 
\be\label{eq:decay}
J_{i_1..i_k}\propto \exp(-{\rm max}|i_\alpha-i_\beta|/\xi_1), 
\ee
where we put lattice spacing to one, and $\xi_1$ is a characteristic length scale (note that this length scale may be different from the localization length defined in terms of the range of LIOMs operators, which will be introduced below). The representation of the Hamiltonian (\ref{eq:Htau},\ref{eq:decay}) implies that the dynamics in the MBL phase is limited to an exponentially slow dephasing between remote pseudospins. It was argued~\cite{Serbyn13-1,Serbyn13-2,HuseOganesyan13} (see also Ref.~\cite{Vosk13} for an alternative, strong-disorder renormalization group approach) that this dephasing underlies the generation, logarithmic in time, of entanglement entropy~\cite{Moore12,Prosen08} for a broad class of initial states.  This reflects the fact that information propagation in the MBL phase is very different from conventional ergodic many-body systems: MBL systems satisfy~\cite{Kim2014} zero-velocity Lieb-Robinson bound~\cite{LiebRobinson}, as opposed to finite-velocity entanglement spreading in ergodic systems~\cite{KimHuse}.

\section{Constructing local integrals of motion}

In this Section, we describe the construction of LIOMs. One straightforward approach would be to attempt constructing the {\it quasi-local} unitary $U$ that diagonalizes the MBL Hamiltonian, and then obtain the complete set of $\tau_i^z$ operators using (\ref{eq:tau}). While in principle possible, practically finding a maximally local representation for the unitary diagonalizing the Hamiltonian turns out to be a challenging task, especially at moderate disorder~\cite{unpublished}. 

Therefore, we take an alternative approach, inspired by the expectation that a local perturbation affects the MBL phase only locally. The key idea is to start with a local physical operator ${\mathcal O}$ (that is, acting on a finite number of local degrees of freedom, e.g., one spin), and evolve it with the MBL Hamiltonian: ${\mathcal O}(t)=e^{iHt} {\mathcal O} e^{-iHt}$. Although transport of energy and spin is absent, the MBL Hamiltonian does generate long-range entanglement (over an exponentially long time)~\cite{Moore12,Serbyn13-1,Serbyn13-2,Vosk13,HuseOganesyan13}, thus, as time $t$ grows, the operator ${\mathcal O}(t)$ becomes more and more nonlocal. However, as we now show, long-time averaging turns it into a local operator, because the non-local terms are oscillating in time. The time-averaged operator is given by: 
\begin{align}
\bar{\mathcal{O}} = \lim_{T\rightarrow \infty} \frac{1}{T}\int_0^T dt \, \mathcal{O}(t) 
\end{align}
The operator $\mathcal{\bar{O}}$ is clearly an IOM because it is diagonal in the energy eigenbasis:
\begin{align}
\label{Eq:OEbasis}
\bar{\mathcal{O}}=\sum_{I}^{2^N}  \la I | {\mathcal O} | I\ra  |I\ra \la I|,
\end{align}
where $I$ runs over all energy eigenstates. We now show that it is also local, by expressing $\mathcal{O}$ and $\bar{\mathcal{O}}$ in the basis of $\tau_i^\alpha$ operators.

Being local, operator $\mathcal O$ can be expressed in terms of a finite number of $\sigma_i^\alpha$ operators and their products. Using the relation
$$\sigma_i^\alpha=U^\dagger \tau_i^\alpha U,$$
let us expand $\mathcal O$ in terms of $\tau_j^\alpha$ operators:
\be\label{eq:sigma_expand}
{\mathcal O} =\sum _{\{\vec{i},\vec{\alpha} \}} A_{\{\vec{i},\vec{\alpha} \}} \tau_{i_1}^{\alpha_1} \tau_{i_2}^{\alpha_2} ...\tau_{i_N}^{\alpha_N}, 
\ee
where $\vec{i}=(i_1, i_2,...i_N)$, $\vec{\alpha}=(\alpha_1, \alpha_2,...,\alpha_N)$ and $\alpha_i=0,x,y,z$ ($\tau_i^0$ is an identity operator). Coefficients $A$ satisfy the normalization condition $\sum_{\{\vec{i},\vec{\alpha} \}} |A_{\{\vec{i},\vec{\alpha} \}}|^2={\rm Tr} {\mathcal O}^2/2^N$ (here $\mathcal{O}$ should be viewed as an operator in the $2^N$-dimensional Hilbert space). The quasi-locality of $U$ guarantees that the coefficients $A$ decay as follows:
\be\label{eq:szbardecay}
A_{i_1..i_k}\propto \exp(-{\rm max}(|i_\alpha-i_\beta|, | i_{\mathcal{O}}-i_\alpha|)/\xi_2), 
\ee 
where $i_{\mathcal{O}}$ are drawn from the support of $\mathcal{O}$ and $\xi_2$ is a characteristic length scale.

The time evolution affects different terms in the expansion (\ref{eq:sigma_expand}) differently: the diagonal terms which are products of one or more $\tau_i^z$ ($\alpha_i=0,z$ for all $i_k$, $k=1,..,N$) commute with the Hamiltonian, and therefore they are invariant under time evolution. The off-diagonal terms which involve at least one $\tau^x$ or $\tau^y$ operator, become oscillatory -- in general, the frequency of these oscillations depends on the state of other pseudospins. The oscillatory terms average to zero at long times, while products of $\tau^z$'s retain their values. Thus, time-averaging turns the local operator $\mathcal{O}$ into a local integral of motion. 

The LIOMs constructed in this way are quite different from the $\tau_i^z$ IOMs introduced in the previous Section as they do not satisfy a Pauli algebra. For example, their eigenvalues are not $\pm 1$ and $\bar{\mathcal{O}}$ does not square to the identity. In general, the $2^N$ eigenvalues of $\bar{\mathcal{O}}$ are non-degenerate; thus they uniquely label the eigenstates. However, in practice many eigenvalues are exponentially close to each other (in system size $N$), and we will consider an extensive number of such LIOMs.

The LIOMs (\ref{Eq:OEbasis}), however, have several advantages. Most importantly, as discussed in the next Section, they have a clear physical interpretation and are measurable. Further, LIOMs constructed in this way are uniquely defined, and do not involve an arbitrary choice in labelling eigenstates (as in previously suggested schemes for constructing LIOMs~\cite{Serbyn13-1,HuseOganesyan13}). Finally, LIOMs of this kind can be constructed numerically, and provide a useful tool for characterizing the properties of the MBL phase.

%

\section{Physical interpretation}

We now turn to the physical interpretation and measurement of the LIOMs $\bar{\mathcal{O}}$.
When $\mathcal O$ represents the density of some extensive conserved quantity, we will argue below that $\bar{\mathcal O}$ describes the propagation of that conserved quantity through the infinite temperature ensemble. This interpretation suggests a systematic way to measure the LIOMs. Further, we will use it to define an order parameter for the MBL phase as well as the localization length. 

The model (\ref{eq:XXZ}) has two global conserved quantities: energy and $z$-projection of the total spin, 
$$
S_z=\sum_{i=1}^N \sigma_i^z. 
$$
Let us first take ${\mathcal O}=\sigma_i^z$. 
Consider the infinite-temperature ensemble with the following density matrix:
\be\label{eq:DM_ini}
\rho(0)=2^{-N}\left(1+\sigma_i^z \right) \otimes \mathds{1}_{i'}, 
\ee
where $\mathds{1}_{i'}$ is a unity operator acting on the bare spins $j\neq i$. This density matrix describes a state with magnetization one at site $i$ and zero elsewhere. All other correlation functions are zero. 

 \begin{figure*}[t]
\label{Fig1}
\begin{center}
\includegraphics[width=1.7\columnwidth]{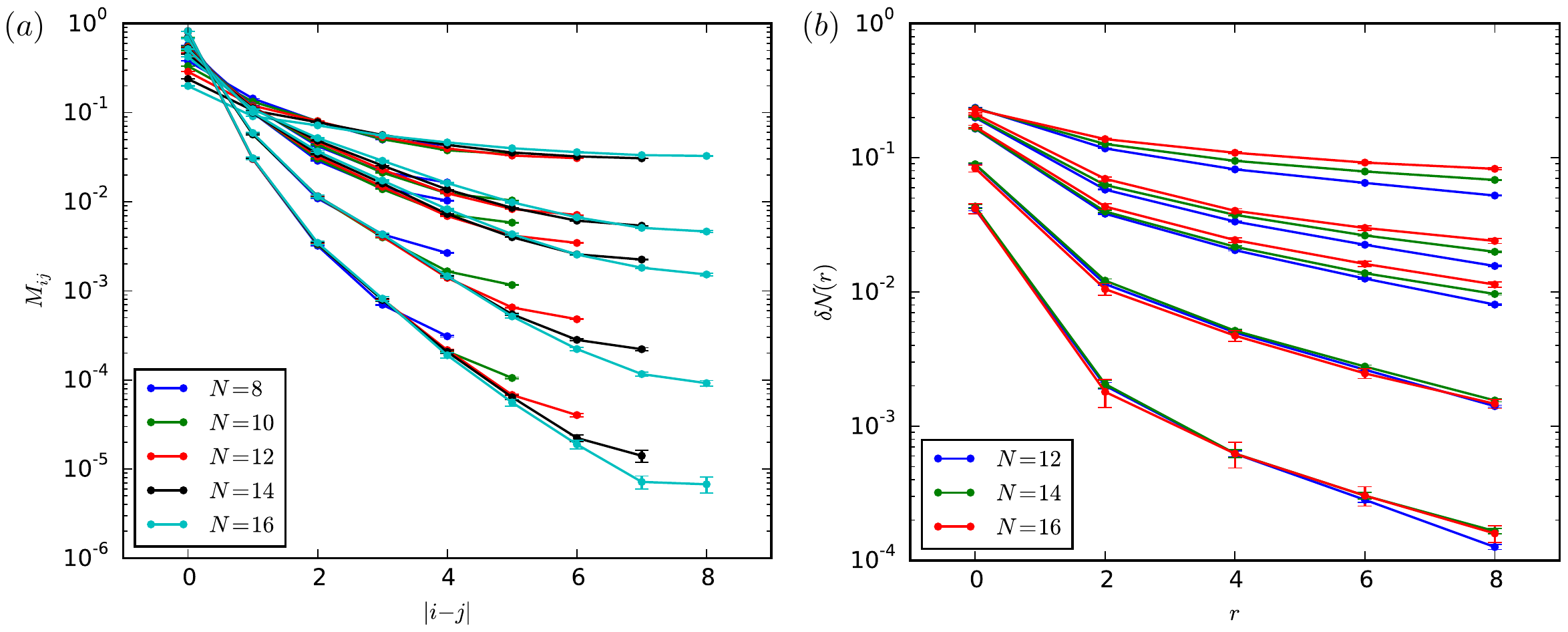} 
\caption{(a) The disorder averaged median $M_{ij}$ vs $|i-j|$ for increasing disorder strengths $W=2,3,3.5,5,7$ from top to bottom for LIOMs $\bar{\sigma}_i^z, i=1\ldots N$ with periodic boundary conditions.
(b) The disorder averaged median of the partial norm $\mathcal{N}(r)$ (Eq.~\eqref{Eq:PartialNorm}) vs the range $r$ of LIOMs $\bar{\sigma}_i^z$ for $i=1, N$ with open boundary conditions. From top to bottom, $W=2,3,3.5,5,7$.
}
\end{center}
\end{figure*}

The up-spin on site $i$ spreads into a region of a finite size with time. Upon time-averaging, the density matrix takes the following form:
\be\label{eq:DM_evo}
\bar\rho=2^{-N}\left(1+\bar{\sigma}_i^z\right), 
\ee
\be\label{eq:sigma_bar}
\bar{\sigma}_i^z=\lim_{T\to \infty} \frac{1}{T} \int_0^{\infty} \sigma_i^z (t) \, dt. 
\ee
Note that the operator $\bar{\sigma}_i^z$ acts non-trivially on all spins, not just on spin $i$.
Measuring different correlation functions in $\bar \rho$ indicates how far the up-spin has propagated from site $i$.
In particular, the time-averaged magnetization $M_{ij}$ on site $j$ is given by: 
\be\label{eq:Mij}
M_{ij}=\frac{{\rm Tr} \left( \bar{\rho}\, \sigma_j^z \right)}{ {\rm Tr} \bar{\rho}}= 2^{-N}{\rm Tr}\left( \bar{\sigma}_i^z \sigma_j^z \right).  
\ee
Here $\bar{\sigma}_i^z$ is considered to be an operator acting in the many-body Hilbert space. 

To better understand the physical meaning of LIOMs and quantities $M_{ij}$, it is instructive to consider the {\it single-particle} localized phase. To be specific, we consider a fermionic 1D tight-binding model with on-site disorder. Such a model is equivalent to the model (\ref{eq:XXZ}) in which $J_z=0$ (the equivalence is established by the Jordan-Wigner transformation). 
We denote the creation/annihilation operator on site $i$ by $a_{i}^\dagger, a_i$, and the creation/annihilation operator of localized eigenstates by $c_i^\dagger, c_i$; operator $c_i^\dagger,c_i$ corresponds to a localized orbital centered near site $i$.  The two sets of operators are related by 
$$
a_i=\sum_j A_{ij} c_j, 
$$ 
where coefficients $A_{ij}$ give amplitudes of the localized wave functions, which decay exponentially with distance $|i-j|$. We choose ${\mathcal O}$ as the density operator on site $i$, ${\mathcal O}=n_i=a_i^\dagger a_i$. Then the corresponding LIOM is given by
$$
\bar{n}_i=\sum_{j} |A_{ij}|^2 c_{j}^\dagger c_j. 
$$
This is a quadratic operator, which characterizes how a particle initially created at site $i$ spreads through the system. In particular, the probability to find the particle at site $k$ at long times, given by
$$
P_{ik}=\sum_{j} |A_{ij}|^2 |A_{jk}|^2, 
$$
decays exponentially with distance $|i-k|$. Coefficient $P_{ii}$ gives the return probability (a generalized participation ratio), which remains finite in the localized phase, and can be used as a diagnostic of localization.  

Quantities $M_{ij}$ introduced above can be viewed as the many-body generalizations of $P_{ij}$. In particular, $M_{ii}$ gives the long-time spin density on the site where spin polarization was initially created. This quantity can be used as an order parameter of the MBL phase. In the MBL phase, it is expected to stay greater than zero as the system size $N$ is taken to infinity. On the other hand, in the delocalized phase, which satisfies ETH, the spin polarization spreads uniformly over the whole system at long times, and therefore $M_{ii}\approx 1/N$ as $N\to \infty$. 

Measuring higher point correlation functions in $\bar{\rho}$ also provides a systematic way to construct $\bar{\sigma}_i^z$.
To see this, let us expand $\bar{\sigma}_i^z$ in the basis of physical spin operators $\sigma_i^\alpha$ and their products:
\begin{align}
\label{Eq:sigmabarPauliexp}
\bar{\sigma}_i^z = \sum _{\{\vec{j},\vec{\alpha} \}} B_{\{\vec{j},\vec{\alpha} \}} \sigma_{j_1}^{\alpha_1} \sigma_{j_2}^{\alpha_2} ...\sigma_{j_N}^{\alpha_N}, 
\end{align} 
where $\alpha_j=0,x,y,z$. 
The coefficients $B$ are related to the correlation functions of $\bar{\rho}$ (see Eq.~\eqref{eq:DM_evo}):
\begin{align}
\label{Eq:Arho0}
B_{\{\vec{j},\vec{\alpha} \}} &= 2^{-N}\textrm{Tr} (\sigma_{j_1}^{\alpha_2} \sigma_{j_2}^{\alpha_2} ...\sigma_{j_N}^{\alpha_N}  \bar{\sigma}_i^z) \\
&=\textrm{Tr} (\sigma_{j_1}^{\alpha_2} \sigma_{j_2}^{\alpha_2} ...\sigma_{j_N}^{\alpha_N}  \bar{\rho})
\end{align} 
Thus, $B_{k,z}=M_{ik}$ etc. 
Due to the quasi-locality of the operator $\bar\sigma_i^z$, the most important coefficients in the expansion above (as measured by their contribution to the operator norm $\textrm{Tr} (\bar{\sigma}_i^z \bar{\sigma}_i^z)$) come from the terms involving spins in the vicinity of site $i$.
Thus, by measuring correlation functions involving more and more spins further and further away from site $i$, one can systematically approximate $\bar{\sigma}_i^z$ to any desired accuracy.

It is useful to summarize the properties satisfied by $M_{ij}$.
\begin{itemize}
\item As total $S_z$ is conserved, $\sum_j M_{ij}=1$
\item $M_{ij}$ is symmetric: $M_{ij} = M_{ji}$, because $M_{ij}=2^{-N}{\rm Tr}(\bar\sigma_i^z \sigma_j^z)=2^{-N}{\rm Tr}(\bar\sigma_i^z \bar\sigma_j^z)$. 
\item From Eq.~\eqref{Eq:sigmabarPauliexp}, it follows that the operator norm $\textrm{Tr} (\bar{\sigma}_i^z \bar{\sigma}_i^z) = M_{ii} 2^N$.
\item From Eq.~\eqref{Eq:sigmabarPauliexp}, it follows that $\textrm{Tr} (\bar{\sigma}_i^z \bar{\sigma}_i^z)  = 2^N \sum_{\{\vec{j},\vec{\alpha} \}} B_{\{\vec{j},\vec{\alpha} \}}^2$. Using the result above, we obtain the inequality $\sum_j M_{ij}^2 \leq M_{ii}$.
\end{itemize}

The LIOMs also provide a natural physical definition of the localization length. The time-averaged magnetization profile is expected to decay exponentially in the MBL phase: 
\be\label{eq:Mdecay}
M_{ij}\propto \exp(-|i-j|/\xi), 
\ee
and therefore can be used to extract the localization length $\xi$, which diverges as the localization-delocalization phase transition is approached.
In the next Section we numerically confirm this behavior for the random-field XXZ model. 

Although we mostly focused on the case when $\mathcal{O}$ represents the local $S_z$ density, we emphasize that our approach applies to {\it any} local operator $\mathcal{O}$. In particular, it is possible to construct LIOMs describing the propagation of energy, when the system's Hamiltonian is given by the sum of local operators:
\be\label{eq:H_local}
H=\sum_i h_i,
\ee
where $h_i$ is a local many-body operator of range $r$, acting on degrees of freedom situated at $i,i\pm 1,..,i\pm r$ (for the case of XXZ model (\ref{eq:XXZ}), $r=1$). 

We construct a set of $N$ LIOMs $\bar{h}_i$, $i=1,...,N$ by choosing ${\mathcal O}=h_i$. Since the Hamiltonian commutes with itself, $H(t)=H$, it can be rewritten as a sum of the LIOMs $\bar{h}_i$, which are mutually commuting:
\be\label{eq:H_commuting}
H=\sum_i \bar{h}_i, \,\,\, [\bar h_i, \bar h_j]=[H,\bar h_i]=0.  
\ee
In the MBL phase, each term is expected to be quasi-local (exponentially decaying tails). It is natural to expect that the localization length defined using operators $\bar{h}_i$ will be equal to the localization length $\xi$ obtained from operators $\bar{\sigma}^z_i$. We also note that the representation (\ref{eq:H_commuting}) can be used to prove zero-velocity Lieb-Robinson bound for information propagation in the MBL phase~\cite{Kim2014}. 

\section{Constructing local integrals of motion numerically}

In this Section, we explicitly construct an extensive set of LIOMs for the disordered XXZ model (\ref{eq:XXZ}) in the MBL phase. We focus on $\bar{\mathcal{O}}=\bar{\sigma}_i^z$ and show its quasi-locality in the MBL phase. By studying the behaviour of $M_{ii}$ and localization length $\xi$ extracted from $M_{ij}$, we determine the location of the transition and examine how locality of IOMs breaks down when the transition is approached from the MBL side. 

We use exact diagonalization to obtain all the eigenstates $|I\rangle$ in the spectrum and then construct $\bar{\mathcal{O}}$ by projecting it to the energy eigenbasis and keeping only the diagonal part: 
$$
\bar{\mathcal O}=\sum_{I=1}^{2^N} \la I| {\mathcal O}| I\ra |I\ra \la I|, 
$$
We choose the parameters in the Hamiltonian (\ref{eq:XXZ}) to be $J_x=J_z=1$, and present data at system sizes $N=8,10,12,14,16$ with periodic and open boundary conditions. The number of disorder realizations at each system size varies from $1000$ to $10000$. 

 \begin{figure}[htbp]
 \label{Fig2} 
\begin{center}
\includegraphics[width=1\columnwidth]{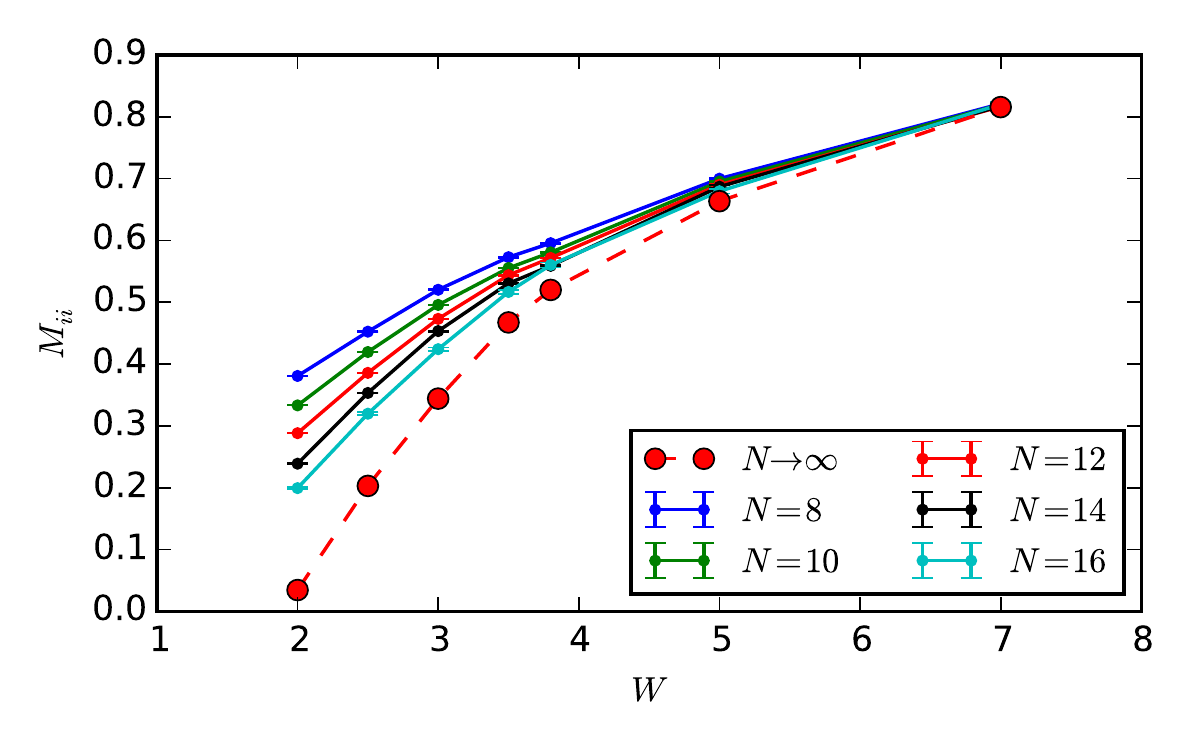}
\caption{ Disorder averaged median of $M_{ii}$ vs disorder strength $W$ at different system sizes. This quantity can be viewed as an order parameter for the MBL phase. The red dashed line shows the linear in $1/N$ extrapolation to the thermodynamic limit.
}
\end{center}
\end{figure}

As expected, in the MBL phase, we find that the LIOMs $\bar{\sigma}^i_z$ are quasi-local, while in the delocalized phase, they are non-local. Fig.1(a) illustrates the disorder averaged median value of $M_{ij}$ vs $|i-j|$ for different disorder strengths $W=2,3,3.5,5,7$. A number of features are worth noting. First, in the MBL phase ($W=3.5,5,7$), the magnetization profile is localized near site $i$: $M_{ii}$ is of the order one, while $M_{ij}$ decays over several orders of magnitude as a function of distance $|i-j|$. Further, there is very little flow with system size. As $W$ decreases, the length scale $\xi$ of decay of $M_{ij}$ increases. In the delocalized phase at $W=2$, $M_{ij}$ is almost independent of $|i-j|$ at the largest system size $N=16$. Further, there is a substantial flow with system size: we have checked that as $N$ increases, $M_{ij}$ approaches the value $1/N$ independent of $|i-j|$, corresponding to the uniform spreading of the spin polarization throughout the system in the ergodic phase. We also note that the systematic higher value of $M_{ij}$ at $|i-j|=N/2$ is due to periodic boundary conditions. 

 To demonstrate the quasi-locality of the entire operator $\bar{\sigma}_i^z$ in the MBL phase, let us compute the partial norm about the site $i$ as a function of a range $r$. Denote the contiguous sites from $i-r$ to $i+r$ by region $A$ and the remaining sites by $\bar{A}$. The number of sites in each region is respectively $N_A$ and $N_B$: $N_A+N_B=N$, $N_A = 2r+1$. The partial norm is defined as:
\begin{align}
\label{Eq:PartialNorm}
\mathcal{N}(r) &= \frac{1}{2^{N_A}}\textrm{Tr} (\bar{\sigma}^A \bar{\sigma}^A ) \\
 \textrm{where } \bar{\sigma}^A &\equiv \frac{1}{2^{N_B}}\textrm{Tr}_{\bar{A}} \bar{\sigma}_i^z
 \end{align}
In other words, $\mathcal{N}(r)$ is the operator 2-norm of the truncated LIOM with support only in region $A$. If the difference between the total norm and the partial norm approaches zero exponentially in $r$, then the operator is quasi-local.

In Fig.1(b), we illustrate the disorder averaged median value of $\delta \mathcal{N}(r)$, the difference between the total norm and the partial norm vs $r$ for the LIOM $\bar{\sigma}_z^i$. To maximize the range accessible at finite size, we use open boundary conditions and restrict our attention to the LIOMs localized at the ends, $i=1, N$.

We argued in the previous Section that the residual magnetization on site, $M_{ii}$, acts as an order parameter for MBL. In Fig.2 the disorder averaged median value of $M_{ii}$ is plotted vs $W$ for different system sizes $N$, as well as the value obtained by $1/N$ extrapolation to the thermodynamic limit. In the MBL phase ($W>4$), the extapolated value of $M_{ii}$ is finite in the thermodynamic limit. Since $M_{ii}$ is nearly independent of $N$ in the MBL phase, the extrapolated value $M_{ii}(\infty)$ is insensitive to the fitting function used. We find that $M_{ii}(\infty)$ becomes zero, indicating a transition into the delocalized phase at $W_*\approx 2$. This value is lower than the previous estimates (e.g., Ref.~\cite{PalHuse} finds that the transition is located at $W_*\approx 3$.) We believe that this discrepancy is due to the linear extrapolation we used, which does not correctly capture the finite-size scaling in the vicinity of the critical point. 

Finally, we extracted the localization length $\xi$ from the spatial decay of magnetization, $M_{ij}\propto \exp(-|i-j|/\xi)$. Due to the finite size effects, the decay length scale depends on the system size, and we obtain the localization length in the thermodynamic limit by linearly extrapolating $1/\xi$ for three largest system sizes ($N=12,14,16$) to $N\to \infty$. The resulting behaviour of $\xi(W)$ is illustrated in Fig. 3. At a critical $W_*\approx 3$ the extrapolated localization length diverges, indicating the transition into a delocalized phase. This critical value is consistent with previous numerical studies~\cite{PalHuse}.

\begin{figure}[t]
\begin{center}
\includegraphics[width=1\columnwidth]{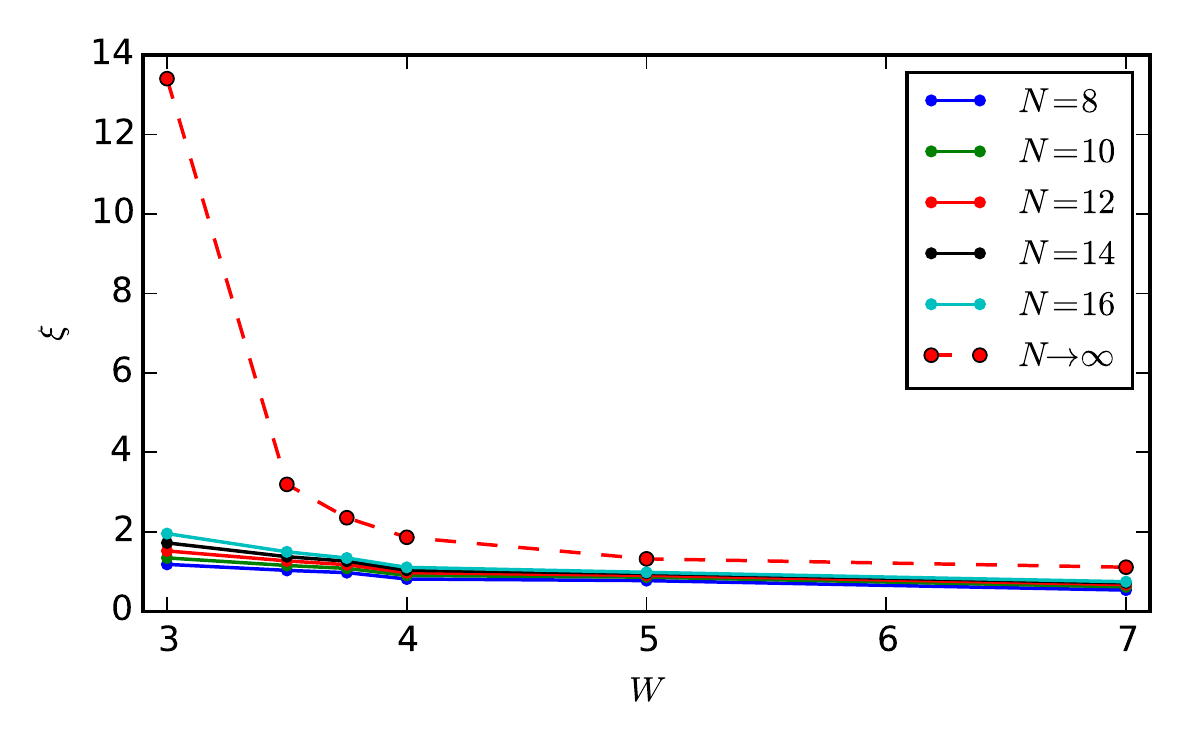}
\label{Fig3}  
\caption{The length scale $\xi$ of decay of $M_{ij}\propto \exp(-|i-j|/\xi)$, for different system sizes. The localization length in the thermodynamic limit (red dashed curve), obtained by a linear extrapolation of $1/\xi$ for three largest system sizes to $N\to \infty$. The extrapolated localization length diverges at a critical $W_*\approx 3$, indicating a transition into the delocalized phase.}
\end{center}
\end{figure}

\section{Conclusions and outlook}

In conclusion, we introduced a new extensive set of local integrals of motion in the many-body localized phase. These LIOMs, obtained by time-evolving and averaging local operators, describe the response of an MBL system to local perturbations. Our approach allowed us to explicitly construct LIOMs in a random-field XXZ spin chain, which exhibits the MBL phase, and explore their properties. We note that the existence of LIOMs in the MBL phase was conjectured previously~\cite{Serbyn13-1,HuseOganesyan13}, but no method for finding them was available. 

The set of LIOMs that we found differs from those considered in previous works in several ways. First, they are uniquely defined, with every local operator generating one integral of motion. This should be contrasted with the approach of Refs.~\cite{Serbyn13-1,HuseOganesyan13}, where LIOMs were (non-uniquely) constructed in terms of projectors onto sets of eigenstates. More importantly, the LIOMs defined above have a clear physical meaning and are experimentally measurable. For cases when the local operator represents a local density of some conserved quantity, the corresponding LIOM describes the propagation of that conserved quantity through the system. The quasi-locality of LIOMs reflects the absence of transport in the MBL phase. Thus, the properties of the LIOMs provide a natural tool for characterizing the MBL phase, both in numerical simulations and in experiments. In particular, such IOMs provide a natural definition of localization length in the MBL phase, and the breakdown of their locality signals a transition into the delocalized phase. 

Experimentally, systems of cold atoms in optical lattices, with their tunable interactions and disorder, appear to be promising candidates for studying many-body localization~\cite{ColdAtoms1,ColdAtoms2,ColdAtoms3}. In particular, no cooling to low temperatures, which is a principal challenge for realizing correlated ground states in these systems, is necessary for seeing MBL. In order to study LIOMs experimentally, in a disordered spin chain, one would prepare a spin-up on one of the sites, and follow the system's evolution. The residual magnetization at long times $M_{ii}$ provides a signature of many-body localization. Moreover, by measuring the magnetization on other sites ($M_{ij}$), as well as correlators of different operators, it would be possible to completely reconstruct the corresponding LIOM and extract its characteristics. We note that Refs.~\cite{Serbyn14-1,Vasseur14} also proposed experiments for probing slow dephasing, which underlies the logarithmic growth of entanglement in the MBL phase.  

Although in this paper we used exact diagonalization to construct LIOMs, the LIOMs defined above can be also obtained using other methods, including matrix-product operator techniques. We also note that there are alternative ways of constructing LIOMs in the MBL phase, which, however, appear to be more challenging to implement. One approach, which will be reported elsewhere~\cite{unpublished}, is to construct the quasi-local unitary $U$ which diagonalizes the Hamiltonian, and then obtain $\tau_i^z$ operators via (\ref{eq:tau}).

Finally, we note that our approach is not limited to conserved quantities: a LIOM can be obtained by time-averaging any local operator, e.g., $\sigma_i^x \sigma_{i+1}^x$. Therefore, our approach can be extended to study many-body localization in systems without global conservation laws (the energy conservation can be broken in periodically driven systems). We expect that in such system in the MBL phase there will also be an extensive set of emergent local integrals of motion. We leave a detailed study of this issue, as well as the existence and structure of LIOMs in MBL systems with a mobility edge, to future work.

\section{Acknowledgements}

Research at Perimeter Institute is supported by the Government of Canada through Industry Canada
and by the Province of Ontario through the Ministry of Economic Development and Innovation. 
We acknowledge support by NSERC Discovery Grant (G.V. and D.A.), the John Templeton Foundation and by the
Simons Foundation (Many Electron Collaboration) (G.V.). We thank Zlatko Papi\'c, Maksym Serbyn, Arijeet Pal, and Chris Laumann for discussions.

%
%
%


\end{document}